\documentclass[12pt,chicago,iop]{emulateapj}
\pdfoutput=0

\usepackage{amsmath,natbib,graphicx}
\usepackage{epsf}
\usepackage{fancyvrb}
\usepackage{epstopdf}
\usepackage{epsfig}
\usepackage{color}
\DeclareGraphicsExtensions{.jpg,.pdf,.png,.eps,.ps}

\begin{document}
\VerbatimFootnotes

\title{THE STEPPENWOLF: A PROPOSAL FOR A HABITABLE PLANET IN INTERSTELLAR SPACE} \shorttitle{The Steppenwolf Planet}


\author{D.~S.~Abbot\altaffilmark{1} and E.~R.~Switzer\altaffilmark{2,3}}

\altaffiltext{1}{Department of the Geophysical Sciences,
University of Chicago,
5734 South Ellis Avenue, Chicago, IL 60637, USA; abbot@uchicago.edu}
\altaffiltext{2}{Kavli Institute for Cosmological Physics,
University of Chicago,
5640 South Ellis Avenue, Chicago, IL 60637, USA; switzer@kicp.uchicago.edu}
\altaffiltext{3}{Department of Astronomy and Astrophysics,
University of Chicago,
5640 South Ellis Avenue, Chicago, IL 60637, USA}

\shortauthors{Abbot \& Switzer}


\begin{abstract}
  Rogue planets have been ejected from their planetary system.  We
  investigate the possibility that a rogue planet could maintain a
  liquid ocean under layers of thermally-insulating water ice and
  frozen gas as a result of geothermal heat flux.  We find that a
  rogue planet of Earth-like composition and age could maintain a
  subglacial liquid ocean if it were $\approx$3.5 times more massive
  than Earth, corresponding to $\approx$8 km of ice. Suppression of
  the melting point by contaminants, a layer of frozen gas, or a
  larger complement of water could significantly reduce the planetary
  mass that is required to maintain a liquid ocean.  Such a planet
  could be detected from reflected solar radiation, and its thermal
  emission could be characterized in the far-IR if it were to pass
  within $\cal{O}$(1000) AU of Earth.
\end{abstract}

\bigskip\bigskip

\section{Introduction}\label{sec:intro}

As a planetary system forms, some planets or planetesimals, referred
to as ``rogue'' planets, can enter hyperbolic orbits and be ejected
from the system as a result of gravitational interactions with gas
giant planets \citep{Lissauer87}.  Furthermore, interaction with
passing stars can eject planets from mature systems
\citep{Laughlin:2000p2278}.  The ability of a rogue planet to support
life is of interest as a sort of pathological example of planetary
habitability, because such a planet could potentially represent a
viable option for interstellar panspermia \citep{Manterola10}, and
because such a planet could be the closest source of extrasolar life
for exploration by humanity in the distant future. Since some sort of
starting point is required to discuss the issue, a planet is often
defined as habitable if it can sustain liquid water at its surface
\citep{kasting93}. \citet{Stevenson99} argued that if a rogue planet
had an extremely high-pressure hydrogen atmosphere,
pressure-broadening of far-infrared absorption by molecular hydrogen
could support liquid water on the planet's surface as a result of the
geothermal heat flux alone, making the planet potentially habitable.
\citet{Debes07} showed that terrestrial planets can be ejected with
moons and that the resulting tidal dissipation could increase the
geothermal heat flux by up to two orders of magnitude for
$\cal{O}$($10^8$ yr).

Subglacial liquid water oceans sustained by internal heat flux on icy
bodies represent an alternative type of habitat.  It is well known
that subglacial oceans are possible on moons around giant planets and
on trans-Neptunian objects in the solar system
\citep{2006Icar..185..258H}, as well as water-rich exoplanets in
distant orbits \citep{2006ApJ...651..535E, 2010ApJ...708.1326F}. A
possible terrestrial analog is Lake Vostok, a $\approx$125 m deep lake
which is sustained by geothermal heat flux under $\approx$4 km of ice
on Antarctica \citep{Kapitsa:1996p2413}. \citet{Laughlin:2000p2278}
have even argued that a terrestrial rogue planet, not attached to any
star and receiving negligible energy at its surface, could sustain a
subglacial liquid ocean if it had a thick enough ice layer. We wish to
consider this point in more depth, including issues such as the
potential for solid-state convection of ice, the potential effect of a
thermally insulating frozen gas layer from outgassing of the mantle on
an Earth-like rogue planet, the effects of melting point suppression
due to contaminants, and observational prospects. By Earth-like, we
mean specifically within an order of magnitude in mass and water
complement, similar in composition of radionuclides in the mantle, and
of similar age.  A subglacial ocean on a rogue planet is interesting
because it could serve as a habitat for life which could, for example,
survive by exploiting chemical energy of rock that is continually
exposed by an active mantle. We will refer to a rogue planet harboring
a subglacial ocean as a Steppenwolf planet, since any life in this
strange habitat would exist like a lone wolf wandering the galactic
steppe.

We can imagine that the ice layer on top of an ocean on a Steppenwolf
planet will grow until either it reaches a steady state with the ice
bottom at the melting point, or all available water freezes. Geothermal
heat from the interior of the Steppenwolf planet will be carried
through the ice layer by conduction, and potentially by convection in
the lower, warmer, and less viscous portion of the ice layer. Since
convection transports heat much more efficiently than conduction, the
steady-state ice thickness will be much larger if convection occurs,
making it harder to maintain a subglacial ocean.



Here we will calculate steady-state ice thicknesses when there is
conduction only and when there is convection in the lower portion of
the ice, and make the conservative assumption that the thicker
solution is valid.  We must acknowledge, however, that it is very
difficult to establish definitively whether convection would occur,
and the resulting ice thickness if it were to occur, without detailed
knowledge of conditions in and microscale composition of the ice
\citep{Barr09}.  More generally, we will make many simplifications,
including considering the question within the framework of a
one-dimensional (vertical) model, since our primary objective is to
establish whether or not a Steppenwolf planet is feasible.

\section{Geophysical Considerations}

First we calculate the conductive steady-state thickness, $H_{\rm cond}$. Above
$\approx$$10$~K, the temperature dependence of the thermal
conductivity of water ice is well-approximated by $k(T)=A T^{-1}$,
where $T$ is the temperature in Kelvin and $A$=651~W~m$^{-1}$
\citep{Petrenko02}. Dimensional analysis shows that thermal
steady state is reached in $\sim 10^6$ years, much shorter than the
timescale of decay of the geothermal heat flux.  Geothermal heat flux
through the shell will be constant at steady state, since no heat is
produced within the ice, as would occur by tidal heating of a frozen
moon. Since the Steppenwolf planets we consider would be much larger
and drier than the icy moons on which subglacial oceans are typically
studied, we can assume that the ice thickness is much less than the
planetary radius, yielding an exponential temperature profile
through the ice and steady-state thickness, so that
\begin{equation}
\label{eq:Hcond}
H_{\rm cond}=\frac{A}{F}\log \left(\frac{T_H}{T_0}\right),
\end{equation}
%
where $T_H$ is the temperature at the ice--water interface (the melting
temperature), $T_0$ is the temperature at the top of the ice, and $F$
is the geothermal heat
flux. 

Decay of radioactive elements in Earth's interior and primordial heat
remaining from Earth's formation lead to an average geothermal heat
flux emanating from Earth's surface of $F_\earth$=0.087~W~m$^{-2}$
\citep{pollack93}.  This heat flux decays with time such that Earth's
geothermal heat flux may have been roughly twice its present value 3~Gyr 
ago \citep{Turcotte80}.  In order to consider Steppenwolf planets
of different sizes, we use the radius--mass scaling $R \propto M^v$ for
super-Earths with $v=0.27$ \citep{valencia06}.  Heuristically, this
yields a geothermal heat flux that scales as $(M/M_\earth)^{1-2v}$, 
or roughly as the square root of the mass.

The pressure at the bottom of the ice layer is $\approx 9$~MPa for
each kilometer of ice, scaling with mass as $(M/M_\earth)^{1-2v}$.
The melting point of pure ice is $250$--$270$~K at pressures less than
$620$~MPa \citep{Choukroun07}, although contaminants to pure ice such
as chloride salts or ammonia could suppress the melting point by
$\approx 50--100$~K \citep{KARGEL:1991p2372,KARGEL:1992p2373,
  2010SSRv..153..185F, 1996Icar..123..101G}.  The steady-state
thickness in the conductive regime is only logarithmically sensitive
to $T_H$, so we will take $T_H=260$~K in the estimates here, except
that we will consider the eutectic point of an ammonia-water ice
mixture at $176$~K to demonstrate the impact of contaminants.  In
steady-state, the temperature at the surface of a Steppenwolf planet
($T_s$), i.e., the top of the ice or frozen gas layer, will be set by
a balance between thermal emission and geothermal heat flux, $F=\sigma
T_s^4$, where $\sigma$ is the Stefan--Boltzmann constant.
Astrophysical radiation backgrounds \citep{mathis83, dole06} are
negligible.

Any gas present in the atmosphere at planetary ejection or outgassed
subsequently by geological processes will tend to freeze into a
low-thermal-conductivity blanket that could allow $T_0$ to exceed
$T_s$. A blanket formed by freezing Earth's current atmosphere would
only increase $T_0$ by about 4~K \citep{Laughlin:2000p2278}. Isostatic
adjustment \citep{Fowler90}, however, will tend to cause some
continents or islands to rise above the layers of water and ice on an
Earth-like Steppenwolf planet with an active mantle, allowing
volcanoes to continuously emit gasses that can freeze onto the water
ice surface.  Here, we will consider carbon dioxide because it is
likely to be outgassed in significant quantities from an Earth-like
planet and it supports a stable layer with relatively high base
temperature relative to other common gases.  To find an upper bound on
the impact of a frozen gas layer, we will assume that the layer is
Rayleigh--Taylor stable with respect to the underlying water--ice, so
it remains as a surface blanket.

To find the temperature at the carbon-dioxide layer base, we note that
the thermal conductivity of carbon dioxide again scales as $T^{-1}$
(here, in a more limited regime), but with constant of proportionality
$A \approx 100$~W~m$^{-1}$ \citep{2003LTP....29..449S}.  We find that
the maximum temperature supported is robustly $\approx 220$~K for
Earth-like Steppenwolf planets and is determined by the weak
temperature dependence of the melting curve
\citep{2006JChPh.125e4504G}. Setting $T_0 = 220$~K reduces the
required steady-state water-ice thickness by an order of magnitude.

At Earth mass, the temperature at the bottom of a layer of solid
CO$_2$ reaches the melting temperature of CO$_2$ for a layer thickness
of $\approx 2$~km, or $\approx 3 \times 10^6$~kg~m$^{-2}$. Venus'
atmosphere has a partial pressure of CO$_2$ of $\approx$90~bar
($\approx 10^6$~kg~m$^{-2}$), which is roughly the vapor pressure of
carbonate rocks at Venus' surface temperature, implying that there may
be more carbon locked in rock in equilibrium with the atmosphere
\citep{Pierrehumbert10}. The store of carbon in carbonate rocks in
Earth's interior is uncertain, but the continental crust is estimated
to contain the equivalent of $\approx 7 \times 10^5$~kg~m$^{-2}$
CO$_2$ and the mantle may contain 2--4 times this amount
\citep{ZHANG:1993p2414}. Therefore it appears reasonable to assume
that a Steppenwolf planet could have a sufficient complement of CO$_2$
to significantly elevate $T_0$.  

Since the viscosity of ice depends strongly on temperature
\citep{Barr09}, if ice convection were to occur on a Steppenwolf
planet, it would occur only in the lower, warmer ice regions and would
be capped by a ``stagnant'' conducting lid \citep{solomatov95}. We
calculate the steady-state ice thickness when convection occurs
following \citet{2006Icar..185..258H}, who assume a Newtonian
rheology. We outline the solution here, but the reader should consult
\citet{2006Icar..185..258H} for more detail.  We assume that
convection occurs below temperature $T_c$ (at higher temperature) and
determine $T_c$ by assuming that the viscosity is reduced by a factor
$\gamma=10$ over the convecting region, where the viscosity is given
by the relation, $\eta(T)=\eta_0 \exp[l(T_m/T-1)]$, where
$\eta_0=10^{13}$~Pa~s, $T_m$ is the melting temperature, and
$l=25$. Assuming that $T_m=T_H$, we can solve for $T_c$.  A
Nusselt--Rayleigh number scaling (Nu $=a$Ra$^\beta$) yields the thickness
of the convecting region between temperature $T_c$ and $T_H$,
\begin{equation}
\label{eq:Hconv}
H_{\rm conv}^{1-3\beta}=\frac{a k (T_H-T_c)}{F }
\left[ \frac{g \alpha \rho (T_H-T_c)}{\kappa \eta(\bar{T})} \right]^{\beta},
\end{equation}
where we evaluate the viscosity at the mean temperature of the
convecting layer, $\bar{T}=\frac{1}{2}(T_c+T_H)$, $\kappa = 1.47
\times 10^{-6}$~m$^2$~s$^{-1}$, $\alpha = 1.56 \times
10^{-4}$~K$^{-1}$, $\rho = 917$~kg~m$^{-3}$, $k =
3.3$~W~m$^{-1}$~K$^{-1}$, $a=0.12$, and $\beta=0.3$. Taking $k$ to be
constant in the convecting layer is a reasonable assumption given that
the temperature is nearly constant within it. We add the thickness of
the convective layer given by Equation (\ref{eq:Hconv}) to the
thickness of the stagnant lid, $H_{\rm lid}=\frac{A}{F}\log
\left(\frac{T_c}{T_0}\right)$, to find the total ice thickness when
there is convection. When the total thickness with convection exceeds
the conductive thickness given by Equation (\ref{eq:Hcond}), we use it
for the ice thickness. This corresponds to a critical Rayleigh number
of roughly 1000 within the convecting layer.

Given that the ice composition on a Steppenwolf planet is unknown and
the ice material properties under appropriate conditions are
poorly-constrained, our convective calculation should be viewed as a
rough estimate. For example, following \citet{2006Icar..185..258H} and
\citet{2010ApJ...708.1326F}, we have assumed Newtonian ice flow, which
may or may not be realistic \citep{Barr09}. Creep mechanisms with
stress-dependent viscosity, however, should yield results roughly
similar to Newtonian flow \citep{KIRK:1987p2435}. Another source of
uncertainty is the appropriate ice grain size, the typical size of
individual components of polycrystalline ice. In general, larger grain
sizes correspond to higher viscosities, making convection less
likely. The value of, $\eta_0$ that we have used, $10^{13}$~Pa~s,
corresponds to an ice grain size of $\approx$0.1 mm
\citep{Barr09}. For reference, the typical ice grain size on
terrestrial ice sheets is $0.1--10$~mm \citep{Barr09} and
\citet{2010ApJ...708.1326F} assume an ice grain size of 0.2 mm for
their calculations of convection on icy extrasolar planets. Competing
estimates of ice grain size on Europa, however, place it at either
$0.02$--$0.06$~mm or $>40$~mm \citep{Barr09}. Although the appropriate
ice grain size on a Steppenwolf planet is a source of uncertainty, we
have used a relatively small value, which is conservative in that it
makes convection more likely to occur. Finally, the constant $\beta$
can take different values between $0.25-0.33$ depending on the
geometry and boundary conditions \citep{2006Icar..185..258H}. We have
used a fairly high value of $\beta$, which leads to a conservatively
strong scaling of $H_{\rm conv}$ with $M$.

\begin{figure}[h]
\begin{center}
\epsfig{file=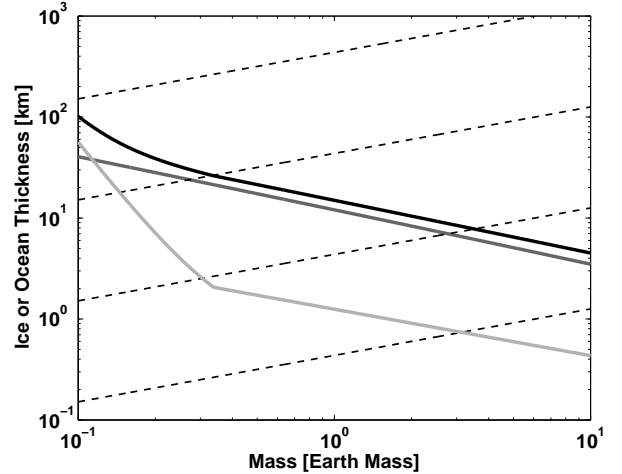, width=8cm} 
\end{center}
\caption{Ice thickness that is required to support liquid water at
  its base as a function of planetary mass, assuming: (1) solid black:
  a melting point of $260$~K and surface temperature required to
  radiate the geothermal heat flux in steady state, (2) dark solid
  gray: a melting point of $176$~K at the eutectic mixture of water
  ammonia \citep{1996Icar..123..101G} and surface temperature
  required to radiate the geothermal heat flux in steady-state, and (3)
  light solid gray: a melting point of $260$~K and water-ice top
  temperature along the melt curve of carbon dioxide at $\approx
  220$~K (providing an upper bound on the possible impact of an
  insulating carbon dioxide layer). The ice thickness is calculated
  using the conductive (Equation (\ref{eq:Hcond})) or stagnant-lid
  convective (Equation (\ref{eq:Hconv})) solution, as
  appropriate. Convection increases the required thickness at low
  mass, and follows the method in \cite{2006Icar..185..258H}.  In the
  case where ammonia has suppressed the melting point by nearly
  $100$~K, we find a modest change in the conductive regime, but that
  convection is suppressed (at fixed water-ice rheology). The dashed
  black lines show the adjusted ocean depth (ocean depth multiplied by
  the ratio of the density of water to that of ice) as a function of
  planetary mass using the simple scaling $M^{1-2 \nu}$ for water
  complements that are 0.1 (lowest line), 1, 10, and 100 (highest
  line) times that of Earth, which is taken to be $4$~km at
  $1~M_\Earth$. A subglacial liquid ocean is possible when the
  adjusted ocean depth exceeds the ice thickness.}
\label{fig:cond_phase}
\end{figure}

A Steppenwolf planet with mean ocean depth greater than the
steady-state ice thickness, accounting for the ice-water density
difference, will have an ocean under its ice layer. Figure
\ref{fig:cond_phase} shows the required ice thickness to achieve
liquid water in several scenarios.  It is expected that the abundance
of water on planetary surfaces varies greatly \citep{Raymond:2007},
but we can consider a simple scaling to understand how the conditions
for liquid water scale with planet mass and water complement.  At
fixed planetary water mass fraction and fraction of water at the
surface (rather than in the mantle), the depth of the ocean scales
approximately as the typical depth at Earth mass ($D^\ast$) times
$(M/M_\earth)^{1-2v}$.  Figure \ref{fig:cond_phase} shows contours for
planets with water complements that are $0.1$, $1$, $10$ and $100$
times that of Earth.  Combining this scaling with Equation
(\ref{eq:Hcond} and \ref{eq:Hconv}), we find that if a Steppenwolf
planet is similar to Earth in water mass fraction ($D^\ast = D_\earth
\approx 4$~km), radionuclide composition, age, and has no frozen
CO$_2$ layer, it must be $\approx 3.5$ times more massive than Earth
to sustain a subglacial liquid ocean. Contaminants which suppress the
melting point have little effect on the conductive ice thickness, but
significantly reduce the ability of the ice to convect. If a
Steppenwolf planet has ten times more water ($D^\ast=10 D_\earth$)
than Earth or if it has a thick frozen CO$_2$ layer which reaches the
maximum temperature of $\approx 220$~K at its base, the planet must be
only $\approx 0.3$ times Earth's mass to have a liquid ocean.

\section{Observational prospects}

We expect that detection of reflected sunlight in the optical
wavebands and IR follow-up present the only viable observational
choices in the near term.  For a single-visit limiting magnitude
$r\approx24.7$ of LSST \citep{Jones09} (and comparable $r=24$ in the
nearer-term Pan-STARRS \citep{2003EM&P...92..465J}), and albedo of
$0.5$, the limiting distance out to which an object can be detected
with reflected sunlight is $\approx 830 (r/R_\earth)^{1/2}$~AU.  The Palomar
survey of $\sim$12,000~${\rm deg}^2$ to magnitude $21.3$ \citep{2009ApJ...694L..45S} has 
discovered no such objects outside of the smaller trans-Neptunians such 
as Sedna.

The baseline requirement to identify a Steppenwolf planet is a
detection of thermal emission in the far-IR.  The flux at the Wien
maximum is
\begin{eqnarray}
  S_{\rm max} &=& (108~{\rm mJy}) \cdot \left ( \frac{T_{\rm s}}{\rm 1~K} \right )^3 \left ( \frac{R}{R_\earth} \right )^2 \left ( \frac{d}{\rm 1~AU} \right )^{-2},
\end{eqnarray}
where $d$ is the Earth-object separation.  At 10~$M_\earth$,
$T_s=46$~K so that $\lambda_{\rm max} =110~\mu{\rm m}$.  Here, the
$Herschel$ PACS instrument reaches a 40 beams/source confusion limit
at a flux of $\approx$$2$~mJy (\citet{2010A&A...518L..30B}, suggesting
a limiting distance of $\approx$$4000$~AU; PACS reaches $10$~mJy at
$5\sigma$ in $1$~hr \citep{Poglitsch10}).  Higher resolution is
required to progress to lower flux limits.  At $200~\mu{\rm m}$, the
planned $25$~m Cornell-Caltech Atacama Telescope (CCAT) would reach
the source confusion
limit\footnote{http://www.submm.org/doc/2006-01-ccat-feasibility.pdf,
  at 30 beams/source.} at $0.36$~mJy.

Photometric microlensing has also been proposed as a method to detect
rogue planets throughout the galaxy \citep{2004ApJ...604..372H}.  If each stellar system ejects
one $M_\Earth$ planet, a survey like the Galactic Exoplanet Survey
Telescope could anticipate $\sim$$20$ detections of rogue
planets \citep{2002ApJ...574..985B, 2010arXiv1012.4486B}.  Typical
distances to these objects would exceed the capabilities of follow-up
that could elucidate their nature. Free-floating super-Jupiters have been discovered \citep{2009A&A...506.1169B}, but these represent a different class of objects.



\section{Discussion and Conclusion}

A Steppenwolf planet's lifetime will be limited by the decay of the
geothermal heat flux, which is determined by the half-life of its
stock of radioisotopes ($^{40}$K, $^{238}$U,$^{232}$Th) and by the
decay of its heat of formation. As these decay times are
$\sim$$1--5$~Gyr, its lifetime is comparable to planets in the
traditional habitable zone of main-sequence stars \citep{kasting93}.

If a Steppenwolf planet harbors life, it could have originated in a
more benign era before ejection from the host star. Alternatively,
after ejection, life could originate around hydrothermal vents, which
are a proposed location for the origin of life on Earth
\citep{1985OLEB...15..327B}. If life can originate and survive on a
Steppenwolf planet, it must be truly ubiquitous in the universe.

We have shown that an Earth-like rogue planet drifting through
interstellar space could harbor a subglacial liquid ocean despite its
low emission temperature, and so might be considered habitable. Such
an object could be detected and followed-up using current technology
if it passed within $\cal{O}$(1000 AU) of Earth.

\section{Acknowledgements}
DSA was supported by a TC Chamberlin Fellowship of the University of
Chicago and by the Canadian Institute for Advanced Research.  ERS
acknowledges support by NSF Physics Frontier Center grant PHY-0114422
to the Kavli Institute of Cosmological Physics.  We thank F. Adams,
F. Ciesla, N. Cowan, R. Fu, C. Hirata, P. Kelly, N. Murray, S. Padin,
L. Page, R. Pierrehumbert, F. Richter, D. Valencia, and an anonymous
reviewer for conversations or comments on early versions of this
Letter, and Amory Lovins for posing the question of what Earth's
temperature would be if there were no Sun.


\begin{thebibliography}{40}
\expandafter\ifx\csname natexlab\endcsname\relax\def\natexlab#1{#1}\fi

\bibitem[{{Baross} \& {Hoffman}(1985)}]{1985OLEB...15..327B}
{Baross}, J.~A., \& {Hoffman}, S.~E. 1985, Orig. of Life Evo. of Bio., 15, 327

\bibitem[{Barr \& Showman(2009)}]{Barr09}
Barr, A.~C., \& Showman, A.~P. 2009, in {Europa} (University of Arizona Press),
  405--430

\bibitem[{{Bennett} \& {Rhie}(2002)}]{2002ApJ...574..985B}
{Bennett}, D.~P., \& {Rhie}, S.~H. 2002, \apj, 574, 985

\bibitem[{{Bennett} {et~al.}(2010){Bennett}, {Anderson}, {Beaulieu}, {Bond},
  {Cheng}, {Cook}, {Friedman}, {Gaudi}, {Gould}, {Jenkins}, {Kimble}, {Lin},
  {Mather}, {Rich}, {Sahu}, {Shao}, {Sumi}, {Tenerelli}, {Udalski}, \&
  {Yock}}]{2010arXiv1012.4486B}
{Bennett}, D.~P., {et~al.} 2010, submitted (arXiv:1012.4486)

\bibitem[{{Berta} {et~al.}(2010){Berta}, {Magnelli}, {Lutz}, {Altieri},
  {Aussel}, {Andreani}, {Bauer}, {Bongiovanni}, {Cava}, {Cepa}, {Cimatti},
  {Daddi}, {Dominguez}, {Elbaz}, {Feuchtgruber}, {F{\"o}rster Schreiber},
  {Genzel}, {Gruppioni}, {Katterloher}, {Magdis}, {Maiolino}, {Nordon},
  {P{\'e}rez Garc{\'{\i}}a}, {Poglitsch}, {Popesso}, {Pozzi}, {Riguccini},
  {Rodighiero}, {Saintonge}, {Santini}, {Sanchez-Portal}, {Shao}, {Sturm},
  {Tacconi}, {Valtchanov}, {Wetzstein}, \& {Wieprecht}}]{2010A&A...518L..30B}
{Berta}, S., {et~al.} 2010, \aap, 518, L30

\bibitem[{{Bihain} {et~al.}(2009){Bihain}, {Rebolo}, {Zapatero Osorio},
  {B{\'e}jar}, {Vill{\'o}-P{\'e}rez}, {D{\'{\i}}az-S{\'a}nchez},
  {P{\'e}rez-Garrido}, {Caballero}, {Bailer-Jones}, {Barrado y Navascu{\'e}s},
  {Eisl{\"o}ffel}, {Forveille}, {Goldman}, {Henning}, {Mart{\'{\i}}n}, \&
  {Mundt}}]{2009A&A...506.1169B}
{Bihain}, G., {et~al.} 2009, \aap, 506, 1169

\bibitem[{Choukroun \& Grasset(2007)}]{Choukroun07}
Choukroun, M., \& Grasset, O. 2007, J. Chem. Phys., 127, 124506

\bibitem[{Debes \& Sigurdsson(2007)}]{Debes07}
Debes, J.~H., \& Sigurdsson, S. 2007, ApJ, 668, L167

\bibitem[{{Dole} {et~al.}(2006){Dole}, {Lagache}, {Puget}, {Caputi},
  {Fern{\'a}ndez-Conde}, {Le Floc'h}, {Papovich}, {P{\'e}rez-Gonz{\'a}lez},
  {Rieke}, \& {Blaylock}}]{dole06}
{Dole}, H., {et~al.} 2006, \aap, 451, 417

\bibitem[{Durand-Manterola(2010)}]{Manterola10}
Durand-Manterola, H.~J. 2010, submitted (arXiv:1010.2735)

\bibitem[{{Ehrenreich} {et~al.}(2006){Ehrenreich}, {Lecavelier des Etangs},
  {Beaulieu}, \& {Grasset}}]{2006ApJ...651..535E}
{Ehrenreich}, D., {Lecavelier des Etangs}, A., {Beaulieu}, J., \& {Grasset}, O.
  2006, ApJ, 651, 535

\bibitem[{{Fortes} \& {Choukroun}(2010)}]{2010SSRv..153..185F}
{Fortes}, A.~D., \& {Choukroun}, M. 2010, \ssr, 153, 185

\bibitem[{Fowler(1990)}]{Fowler90}
Fowler, C. 1990, The Solid Earth: An Introduction to Global Geophysics
  (Cambridge University Press), 472

\bibitem[{{Fu} {et~al.}(2010){Fu}, {O'Connell}, \&
  {Sasselov}}]{2010ApJ...708.1326F}
{Fu}, R., {O'Connell}, R.~J., \& {Sasselov}, D.~D. 2010, ApJ, 708, 1326

\bibitem[{{Giordano} {et~al.}(2006){Giordano}, {Datchi}, \&
  {Dewaele}}]{2006JChPh.125e4504G}
{Giordano}, V.~M., {Datchi}, F., \& {Dewaele}, A. 2006, J. Chem. Phys., 125,
  054504

\bibitem[{{Grasset} \& {Sotin}(1996)}]{1996Icar..123..101G}
{Grasset}, O., \& {Sotin}, C. 1996, Icarus, 123, 101

\bibitem[{{Han} {et~al.}(2004){Han}, {Chung}, {Kim}, {Park}, {Ryu}, {Kang}, \&
  {Lee}}]{2004ApJ...604..372H}
{Han}, C., {Chung}, S., {Kim}, D., {Park}, B., {Ryu}, Y., {Kang}, S., \& {Lee},
  D.~W. 2004, ApJ, 604, 372

\bibitem[{{Hussmann} {et~al.}(2006){Hussmann}, {Sohl}, \&
  {Spohn}}]{2006Icar..185..258H}
{Hussmann}, H., {Sohl}, F., \& {Spohn}, T. 2006, Icarus, 185, 258

\bibitem[{{Jewitt}(2003)}]{2003EM&P...92..465J}
{Jewitt}, D. 2003, Earth, Moon, Planets, 92, 465

\bibitem[{{Jones} {et~al.}(2009){Jones}, {Chesley}, {Connolly}, {Harris},
  {Ivezic}, {Knezevic}, {Kubica}, {Milani}, \& {Trilling}}]{Jones09}
{Jones}, R.~L., {et~al.} 2009, Earth, Moon, and Planets, 105, 101

\bibitem[{Kapitsa {et~al.}(1996)Kapitsa, Ridley, Robin, Siegert, \&
  Zotikov}]{Kapitsa:1996p2413}
Kapitsa, A., Ridley, J., Robin, G., Siegert, M., \& Zotikov, I. 1996, Nature,
  381, 684

\bibitem[{Kargel(1991)}]{KARGEL:1991p2372}
Kargel, J. 1991, Icarus, 94, 368

\bibitem[{Kargel(1992)}]{KARGEL:1992p2373}
---. 1992, Icarus, 100, 556

\bibitem[{Kasting {et~al.}(1993)Kasting, Whitmire, \& Reynolds}]{kasting93}
Kasting, J.~F., Whitmire, D.~P., \& Reynolds, R.~T. 1993, Icarus, 101, 108

\bibitem[{Kirk \& Stevenson(1987)}]{KIRK:1987p2435}
Kirk, R., \& Stevenson, D. 1987, Icarus, 69, 91

\bibitem[{Laughlin \& Adams(2000)}]{Laughlin:2000p2278}
Laughlin, G., \& Adams, F. 2000, Icarus, 145, 614

\bibitem[{Lissauer(1987)}]{Lissauer87}
Lissauer, J. 1987, Icarus, 69, 249

\bibitem[{Mathis {et~al.}(1983)Mathis, Mezger, \& Panagia}]{mathis83}
Mathis, J., Mezger, P., \& Panagia, N. 1983, A\&A, 128, 212

\bibitem[{Petrenko \& Whitworth(2002)}]{Petrenko02}
Petrenko, V.~F., \& Whitworth, R.~W. 2002, Physics of Ice (Oxford Univ. Press),
  392

\bibitem[{Pierrehumbert(2010)}]{Pierrehumbert10}
Pierrehumbert, R.~T. 2010, Principles of Planetary Climate (Cambridge
  University Press), 652

\bibitem[{{Poglitsch} {et~al.}(2010){Poglitsch}, {Waelkens}, {Geis},
  {Feuchtgruber}, {Vandenbussche}, {Rodriguez}, {Krause}, {Renotte}, {van
  Hoof}, {Saraceno}, {Cepa}, {Kerschbaum}, {Agn{\`e}se}, {Ali}, {Altieri},
  {Andreani}, {Augueres}, {Balog}, {Barl}, {Bauer}, {Belbachir}, {Benedettini},
  {Billot}, {Boulade}, {Bischof}, {Blommaert}, {Callut}, {Cara}, {Cerulli},
  {Cesarsky}, {Contursi}, {Creten}, {De Meester}, {Doublier}, {Doumayrou},
  {Duband}, {Exter}, {Genzel}, {Gillis}, {Gr{\"o}zinger}, {Henning},
  {Herreros}, {Huygen}, {Inguscio}, {Jakob}, {Jamar}, {Jean}, {de Jong},
  {Katterloher}, {Kiss}, {Klaas}, {Lemke}, {Lutz}, {Madden}, {Marquet},
  {Martignac}, {Mazy}, {Merken}, {Montfort}, {Morbidelli}, {M{\"u}ller},
  {Nielbock}, {Okumura}, {Orfei}, {Ottensamer}, {Pezzuto}, {Popesso},
  {Putzeys}, {Regibo}, {Reveret}, {Royer}, {Sauvage}, {Schreiber}, {Stegmaier},
  {Schmitt}, {Schubert}, {Sturm}, {Thiel}, {Tofani}, {Vavrek}, {Wetzstein},
  {Wieprecht}, \& {Wiezorrek}}]{Poglitsch10}
{Poglitsch}, A., {et~al.} 2010, \aap, 518, L2

\bibitem[{Pollack {et~al.}(1993)Pollack, Hurter, \& Johnson}]{pollack93}
Pollack, H., Hurter, S., \& Johnson, J. 1993, Rev. Geophys., 31, 267

\bibitem[{Raymond {et~al.}(2007)Raymond, Scalo, \& Meadows}]{Raymond:2007}
Raymond, S.~N., Scalo, J., \& Meadows, V.~S. 2007, ApJ, 669, 606

\bibitem[{{Schwamb} {et~al.}(2009){Schwamb}, {Brown}, \&
  {Rabinowitz}}]{2009ApJ...694L..45S}
{Schwamb}, M.~E., {Brown}, M.~E., \& {Rabinowitz}, D.~L. 2009, ApJ, 694, L45

\bibitem[{Solomatov(1995)}]{solomatov95}
Solomatov, V. 1995, Phys. Fluids, 7, 266

\bibitem[{{Stevenson}(1999)}]{Stevenson99}
{Stevenson}, D.~J. 1999, Nature, 400, 32

\bibitem[{{Sumarokov} {et~al.}(2003){Sumarokov}, {Stachowiak}, \&
  {Je{\.z}owski}}]{2003LTP....29..449S}
{Sumarokov}, V.~V., {Stachowiak}, P., \& {Je{\.z}owski}, A. 2003, Low Temp.
  Phys., 29, 449

\bibitem[{Turcotte(1980)}]{Turcotte80}
Turcotte, D. 1980, Earth Planet. Sci. Lett., 48, 53

\bibitem[{Valencia {et~al.}(2006)Valencia, O'Connell, \& Sasselov}]{valencia06}
Valencia, D., O'Connell, R., \& Sasselov, D. 2006, Icarus, 181, 545

\bibitem[{Zhang \& Zindler(1993)}]{ZHANG:1993p2414}
Zhang, Y., \& Zindler, A. 1993, Earth Planet. Sci. Lett., 117, 331

\end{thebibliography}

\end{document}